\documentclass[12pt,a4]{article}
\begin{document}

\begin{center}
{\Large \bf The Painleve  Analysis  and
 Special Solutions for Nonintegrable Systems}\\  

\vspace{4mm}

S.Yu.~Vernov\footnote{E-mail:~svernov@theory.sinp.msu.ru, \ home page:
http://theory.sinp.msu.ru/$\sim$svernov}\\[4mm]
{\it Skobeltsyn
Institute of Nuclear Physics\\ Moscow State University,}\\ 
Vorob'evy Gory,  Moscow, 119992, Russia\\
\end{center}

\begin{abstract}
{\normalsize The H\'enon--Heiles system in the general
form is studied. In a nonintegrable case new solutions
have been found as formal Laurent series, depending on three parameters. 
One of parameters determines a location of the singularity point, other 
parameters determine coefficients of the Laurent series.  For some values 
of these two parameters the obtained Laurent series coincide with 
the Laurent series of the known exact solutions.} 
\end{abstract}

\section{The H\'enon--Heiles Hamiltonian}

Let us consider a three-dimensional galaxy  
with an axial-symmetric and time-independent potential function.
The equations of galactic motion admit two well-known 
integrals: energy and angular momentum. If we know also the third integral 
of motion, then we can solve the motion equations by the method of 
quadratures. Due to the symmetry of the potential the considered system
is equivalent to two-dimensional one. However, for many 
polynomial potentials the obtained system has not
the second integral as a polynomial function. 

In the 1960s numerical~\cite{HeHe} and asymptotic 
methods~\cite{Contop,Gustavson} have been developed 
to show either existence or absence of the third integral for some 
polynomial potentials.  To answer the question about the existence of 
the third integral \ H\'enon \ and  \  Heiles~\cite{HeHe} \  considered 
the behavior of numerically integrated trajectories.  They 
wrote~\cite{HeHe}:  "In order to have more freedom of experimentation, we 
 forgot momentarily the astronomical origin of the problem and consider 
its general form: does an axisymmetrical potential admit a third isolating 
integral of motion~?".  They have proposed the following Hamiltonian:  
$$ 
H=\frac{1}{2}\Big(x_t^2+y_t^2+ 
x^2+y^2\Big)+x^2y-\frac{1}{3}y^3,\eqno(1)
$$
because: { \bf (a) } \ it is
analytically simple; this makes the numerical computations of trajectories
easy; { \bf (b) } \ at the  same time, it is sufficiently complicated to
 give trajectories, which are far from trivial.  Indeed, H\'enon and Heiles
found that for low energies this system appeared to be integrable, in so
much as trajectories (numerically integrated) always lay on well-defined
two-dimensional surfaces. On the other hand, they also obtained that for
high energies many of these integral surfaces were destroyed and that
phase space acquired large ergodic regions. The H\'enon--Heiles system
became a paradigm of chaotic Hamiltonian  dynamics. 

Let us consider the { \bf  H\'enon--Heiles } Hamiltonian in the general
form:
$$
H=\frac{1}{2}\Big(x_t^2+y_t^2+\lambda
x^2+y^2\Big)+x^2y-\frac{C}{3}y^3, \eqno(1)
$$
where $C$ and $\lambda$ are numerical parameters.

Investigations of the generalized H\'enon--Heiles system:
$$
  \left\{ \begin{array}{lcl} x_{tt} & = &-\lambda x -2xy,\\[2mm]
  y_{tt} & = &-y -x^2+Cy^2,
\end{array}
\right.
\eqno(2)
$$
 can be separated on the following ways:

{\bf 1. } Numerical analysis~\cite{ChTW1,ChTW2} 
 showed, that in the original case ($\lambda=1$, $C=1$) 
singular points of solutions of the motion equations group in 
self-similar spirals. It turns out extremely difficult distributions of 
singularities, forming a boundary, across which the solutions can not be 
analytically continued. 
Numerical investigations of the generalized H\'enon--Heiles system are 
continued up to now~\cite{Brasil1,Brasil2}.

{\bf 2. } The procedure for transformation the Hamiltonian to a normal form 
and for construction the second independent integral in the form of
formal power series in the phase variables $x$, $x_t$, $y$ ¨ $y_t$ 
(Gustavson integral) has been realized for 
the H\'enon--Heiles system both in the original ($\lambda=1$,
$C=1$)~\cite{Gustavson} (see also~\cite{Moser}) and in the general
forms~\cite{Braun}. Using the Bruno algorithm~\cite{Bruno1,Bruno2} 
\ V.F.~Edneral \  has constructed the Poincar\'e--Dulac normal form and 
found  
(provided that all phase variables are small) local families of 
periodic solutions of the H\'enon--Heiles
system both in the original~\cite{Edneral1} and in the
general~\cite{Edneral2} forms.

 {\bf 3. }  The singularities at the fixed points in phase space are
locally analyzed via normal form theory, whereas the singularities in
the complex (time) plane are studied by the Painlev\'e analysis. Using
this analysis three integrable cases of the generalized H\'enon--Heiles system
have been found:
$$
\begin{array}{cll}
 \mbox{(i)} & C=-1, &\lambda=1,\\
 \mbox{(ii)} & C=-6, &\lambda \mbox{\ \bf is an arbitrary number},\\
 \mbox{(iii)} & C=-16,\quad &\lambda=\frac{1}{16}.\\
\end{array}
$$

In contradiction to the case (i) the cases (ii) and (iii)
are nontrivial, so the integrability of these cases had to be proved
additionally. In the 1980's the required second integrals were 
constructed~[14--18].  For integrable cases of the  H\'enon--Heiles system 
the B\"acklund transformations~\cite{Weiss1} and the Lax 
representations~\cite{NTZ,Polska} have been found.  
The three integrable cases of the H\'enon--Heiles system correspond
precisely to the stationary flows of the only three integrable
cases of {\it fifth-order polynomial nonlinear evolution} equations of
scale weight 7 (respectively the Sawada--Kotega, the fifth-order
Korteweg--de Vries and the Kaup--Kupershmidt 
equations)~\cite{Polska,Fordy2}.  The general solutions of the 
H\'enon--Heiles system in integrable cases are known~\cite{Conte4}.

{\bf 4. } 
The H\'enon--Heiles system as a system of two second order ODEs
is equivalent to the fourth order equation\footnote{For given $y(t)$ the 
function $x^2(t)$ is a solution of a linear equation. System $(2)$ is 
invariant to exchange $x$ to $-x$. }:  
$$ 
y_{tttt}=(2C-8)y_{tt}y - 
(4\lambda+1)y_{tt}+2(C+1)y_{t}^2+ \frac{20C}{3}y^3+ 
(4C\lambda-6)y^2-\lambda y-4H. \eqno(3) 
$$

To find a special solution of
this equation one can assume that $y$ satisfies some more simple
equation.  For example, it is well known that the
H\'enon--Heiles system and, hence, equation~$(3)$ have solutions
in terms of the Weierstrass elliptic functions satisfying the
first-order differential equation:
$$
  y_t^2={\cal A}y^3+{\cal B}y^2+{\cal C} y+ {\cal D}, \eqno(4)
$$
where $\cal A$, $\cal B$, $\cal C$ and $\cal D$ are some constants.

E.I. Timoshkova~\cite{Timosh} generalized equation $(4)$:
$$
  y_t^2={\cal  A}y^3+{\cal B} y^2+{\cal C} y+{\cal D}
 +{\cal G} y^{5/2}+ {\cal E} y^{3/2}
\eqno(5)
$$
and found new one-parameter sets of solutions of the 
H\'enon--Heiles  system in nonintegrable cases
($C=-\:\frac{4}{3}$ \ or \ $C=-\:\frac{16}{5}$, \ $\lambda$ is 
an arbitrary number).

In the present paper I use the Painlev\'e method to find  
asymptotic solutions of the H\'enon--Heiles system at 
$C=-\:\frac{16}{5}$.

\section{The Painlev\'e property}

Let us formulate the Painlev\'e property for ODE's. 
Solutions of a system of ODE's  
are regarded as analytic functions, may be 
with isolated singular points~\cite{Golubev1, Hille}. 
A singular point of a solution is said {\it critical } (as opposed to   
{\it noncritical}) if the solution is multivalued (single-valued)
in its neighborhood and {\it  movable} if its location 
depends on initial conditions\footnote{Solutions of a system with a
time-independed Hamiltonian can have only movable singularities.}.

\vspace{2.7mm}

{\large\it Definition. } {\it A system of  ODE's has 
 \textbf{\textit{ the Painlev\'e
property }} if its general solution has no movable 
critical singular point}~\cite{Painleve}.

\vspace{2.7mm}

An arbitrary solution of such system is single-valued in the neighborhood 
of its singular point $t_0$ and can be expressed as a Laurent series with
a finite number of terms with negative powers of $t-t_0$. 

A Hamiltonian system in a
$2s$--dimensional phase space is called { \it completely integrable } if it
possesses $s$ independent integrals which commute with respect
to the associated Poisson bracket. When this is the case, the equations of
motion are (in principal, at least) separable and solutions can be obtained
by the method of quadratures.
 Since the work of
S.V.~Kovalevskaya~\cite{Kova} (see also~\cite{Golubev2}) on the motion of
a heavy rigid body about a fixed point, the Painlev\'e property has been 
proposed as a criterion for complete integrability~[30--32].  
 If the system misses the Painlev\'e property (has complex or irrational 
"resonances"), then the system cannot be "algebraically
integrable"~\cite{Y1} (see also~\cite{Y2} and references there in).
 N.~Ercolani and E.D.~Siggia~\cite{ES1,ES2} advance arguments as
to why the  Painlev\'e test works,  {\it i.e.} they showed how to exploit
the singular analysis to yield the integrals. They proved a theorem which
demonstrates that the singularity analysis provides bounds on the degrees
of polynomial integrals for a large class of separable systems.

\section{The Painlev\'e test}
\subsection{Various algorithms of the Painlev\'e test}
The Painlev\'e test is any  algorithm designed to determine necessary
conditions for a differential equation to have the Painlev\'e property.
The original algorithm developed by Painlev\'e~\cite{Painleve} is known
as the $\alpha$-method.  The method of S.V.~Kovalevskaya is
not as general as the $\alpha$-method but is more simple than it is.  

In 1980, motivated by the work of S.V.~Kovalevskaya~\cite{Kova},
 \  M.J.~Ablowitz, A.~Ramani and H.~Segur~\cite{ARS} 
developed a new algorithm of the
Painlev\'e test for ODE's~\footnote{They also were the first~\cite{ARS,
ARS2} to point out the connection between the nonlinear partial differential 
equations (PDE's), which are soluble by the inverse scattering transform 
method, and ODE's with the Painlev\'e property. They have proven that if 
a PDE is solvable by the inverse 
scattering transform and a system of ODE's is obtained from this PDE by 
the exact similarity reduction then the solutions (of this system of 
ODE's) associated with Gel'fand--Levitan--Marchenko equation will possess 
the Painlev\'e property. Furthermore, they conjecture that, when all the 
ODE's obtained by exact similarity transforms from a given PDE have the 
Painlev\'e property, perhaps after a change of variables, then PDE will be 
"integrable". Subsequently the Painlev\'e property 
for PDE was defined and the corresponding Painlev\'e test (the WTC 
procedure) was constructed by J.Weiss, M.Tabor and 
G.Carnevale~\cite{WTC,Weiss4} (see  also~[41--48]).   
For many integrable PDE's, for example, the Korteweg--de-Vries 
equation~\cite{Tabor}, the B\"acklund transformations and the Lax 
representations result from the WTC procedure~\cite{Weiss4, 
Musette1,ConteTMF}.  Also, special solutions for certain nonintegrable 
PDE's were constructed using this algorithm~\cite{Tabor2,Conte3}.  The 
Painlev\'e test for both ODE's and PDE's which is based on perturbation 
theory is presented in~\cite{Conte2,Conte21}.}.  Using this algorithm one 
  can determine whether a system of ODE's admits movable branch points, 
either algebraic or logarithmic.  This algorithm can be used not only to 
isolate values of parameters of integrable cases, but also to find special 
asymptotic solutions even in nonintegrable cases~\cite{Sahadevan}.
The Painlev\'e test finds wide use in theoretical physics 
as procedure for analysis of complex systems, for example, 
the self-dual Yang-Mills equation~\cite{JKM}
or the Bianchi IX cosmological 
model~[53--56]. 

If one substitutes in a system of ODE's, for example, in system (2):  
$$ 
 x=\widetilde{cx}_{\alpha}(t-t_0)^{\alpha}+\sum\limits_{j=1}^{N_{max}} 
\widetilde{cx}_{j+\alpha}(t-t_0)^{j+\alpha}\quad\mbox{and}\quad
y=\widetilde{cy}_{\beta}^{\vphantom{27}}(t-t_0)^{\beta}+
\sum\limits_{j=1}^{N_{max}}
\widetilde{cy}_{j+\beta}(t-t_0)^{j+\beta},
$$
where $N_{max}$, \ $\alpha$, \ $\beta$, \ 
$\widetilde{cx}_\alpha^{\vphantom{27}}$ 
and $\widetilde{cy}_\beta^{\vphantom{27}}$ are some known constants, 
then the system of ODE's transforms into a set of linear 
algebraic systems in coefficients $\widetilde{cx}_{k}$ and 
$\widetilde{cy}_{k}$. In the general case one can obtain the exact 
solutions (in the form of formal Laurent series) only if one solves 
infinity number of systems: $N_{max}=\infty$. On the other hand, 
if one solves a finite number of systems then one obtains asymptotic 
solutions.  With the help of some computer algebra system, for example, 
the system {\bf REDUCE}~[57--59], these systems can be 
solved step by step and asymptotic solutions can be automatically found 
with any accuracy. But previously one has to determine values of constants 
$\alpha$, $\beta$, $\widetilde{cx}_\alpha^{\vphantom{27}}$ and 
$\widetilde{cy}_\beta^{\vphantom{27}}$ and to analyze
systems with zero determinants. Such systems
correspond to new arbitrary constants or have no solutions.
Powers at which new arbitrary constants enter are called {\it resonances}.
The Painlev\'e test gives all information about possible dominant 
behaviors and resonances. Moreover, the results of the  Painlev\'e 
analysis  prompt cases, in which it is useful to include into expansion 
terms with fractional powers of  $t-t_0$.   

The test consists of three levels, if
the system passes some level then it means that the system may be a system
of P-type and we have to check the system on the following level.

\subsection{The first level: find the dominant behavior}
We assume that the dominant behavior of solutions in a sufficiently small
neighborhood of the singularity is algebraic. To find the dominant  behavior
we look for solutions in the form
$$
x=a_1(t-t_0)^\alpha\qquad \mbox{and}\qquad y=a_2(t-t_0)^\beta,\eqno(6)
$$
where $t_0$ is arbitrary. Substitution~$(6)$ into~$(2)$ shows that for
certain values of $\alpha$ and $\beta$, two or more terms in the equations
of~$(2)$ may balance (these terms have the same powers),
and the rest can be ignored as $t\longrightarrow t_0$. For each choice of
$\alpha$ and $\beta$ the terms which can balance are called {\it the
leading terms}. For the H\'enon--Heiles system in the general
form there exist two possible dominant behaviors~\cite{ChTW2,BSV,Tabor}:

\vspace{5mm}
\begin{tabular}{|l|l|}
\hline
{ \it Case 1}:  &  { \it Case 2}: ($\beta<\Re e(\alpha)<0$)\\[2mm]
\hline
 $\alpha=-2$,$\qquad\qquad\qquad\beta=-2$ &
$\alpha=\frac{1\pm\sqrt{1-48/C}^{\vphantom{7^4}}}{2}$,
 $\qquad\beta=-2$\\[1mm]
$a_1=\pm 3\sqrt{2+C}$,$\qquad a_2=-3$ & $a_1=arbitrary$,
 $\qquad a_2=-\:\frac{6}{C}$ \\[1mm]
  \hline
 \end{tabular}
\vspace{5mm}

It is possible that an original system is not of
P-type, but, after some change of variables, the obtained system is of
P-type\footnote{In this case the original system is said to have the weak
Painlev\'e property. For example, in the integrable case (iii)
system (2) has the weak Painlev\'e property. The interesting example of
a system with the weak Painlev\'e property is
presented in~\cite{Sako2}.}. 
At $C=-\:\frac{16}{5}$ \
in the {\it Case 2} we obtain $\alpha=-\:\frac{3}{2}$ and to continue the
Painlev\'e test we have to introduce new variable $z=x^2$ and to consider
the following system instead of system $(2)$:
$$
\left\{ \begin{array}{ccl} z_{tt}z & = &\frac{1}{2}z_t^2
-2\lambda z^2 -4z^2y,\\[1mm] y_{tt} & = &-y -z-\frac{16}{5}y^2. 
 \end{array} \right.
\eqno(7)
$$

\subsection{The second level: find the resonances}

The second level of the Painlev\'e analysis is finding the resonances.
For each obtained pair of values of $\tilde\alpha\equiv 2\alpha$ and
$\beta$ we construct the simplified system that retains only the leadings
terms of equations of the original system $(7)$.

For $\tilde\alpha=-4$ and $\beta=-2$ the simplified system is
$$
\left\{ \begin{array}{ccl}
z_{tt}z & = &\frac{1}{2}z_t^2
 -4z^2y,\\[2mm]
y_{tt} & = &-z-\frac{16}{5}y^2.
  \end{array}
\right.
\eqno(8.1)
$$

Substituting
$$
z=-\:\frac{54}{5}(t-t_0)^{-4}+b_1(t-t_0)^{-4+r} \quad\mbox{and}\quad
  y=-3(t-t_0)^{-2}+b_2(t-t_0)^{-2+r}
$$
in  system $(8.1)$, we obtain that to leading order in $(t-t_0)$
\{$(t-t_0)^{r-8}$ for the first equation and
  $(t-t_0)^{r-4}$ for the second equation\} this system
reduces to a system of two linear algebraic equations:
$$
  \hat Q(r)\bar b=0,      \eqno  (9)
$$
where  $\hat Q(r)$  is $2\!\!\times\!\!2$ matrix, which elements depend on
$r$, and $\bar b\equiv(b_1,b_2)$.  Determinant $det(\hat Q(r))$ is a
polynomial of order 4. Equation $(9)$ has nonzero solution
only if 
$$
det(\hat Q(r))=0.          \eqno (10)
$$ 
The roots of equation $(10)$:
$$
r_1=-1,\qquad r_2=6,\qquad r_3=\frac{5  +\sqrt{53.8}}{2}, \qquad 
r_4=\frac{5 -\sqrt{53.8}}{2} 
$$ 
determine resonances (one root is always 
$(-1)$, it represents the arbitrariness of $t_0$). Some roots of $(10)$ 
are not integer. This result means, that the  H\'enon--Heiles system with 
$C=-\:\frac{16}{5}$ is a nonintegrable system. There is no algorithm to
find the general solution for a nonintegrable system. 

To find special asymptotic solutions let us consider the dominant
behavior in the {\it Case 2}  ($\tilde\alpha=-3$ and $\beta=-2$).
The simplified  system is\footnote{The simplified systems $(8.1)$ and
$(8.2)$ are different.}
 $$
 \left\{ \begin{array}{ccl} z_{tt}z & = &\frac{1}{2}z_t^2
 -4z^2y,\\[2mm]
 y_{tt} & = &-\:\frac{16}{5}y^2.
  \end{array}
\right.
\eqno(8.2)
$$

Substituting ($\tilde a_1\equiv a_1^2$)
$$
  z=\tilde a_1(t-t_0)^{-3}+b_1(t-t_0)^{-3+r}\quad\mbox{and}\quad
y=-\:\frac{15}{8}(t-t_0)^{-2}+b_2(t-t_0)^{-2+r}
$$
into $(8.2)$ we repeat calculations and obtain that resonances and
corresponding arbitrary parameters can arise in
terms proportional to $(t-t_0)^{r-2}$, where $r=-1,\:0,\:4,\:6$.
Root $r=-1$ corresponds to arbitrary parameter
$t_0$, root $r=0$ corresponds to arbitrary parameter 
$\tilde a_1$, other roots
correspond to new arbitrary parameters, {\it i.e.} new constants of
integration. For arbitrary $C$ values of 
resonances $r$ are~\cite{ChTW2,Tabor}:

\vspace{5mm}
\begin{tabular}{|l|l|}
\hline
{ \it Case 1}:  &  { \it Case 2}: \\[2mm]
\hline
 $\tilde\alpha=-4$, & 
 $\tilde\alpha=\pm\sqrt{1-48/C}^{\vphantom{7^4}}$,\\[1mm]
 $\beta=-2$, &
 $\beta=-2$,\\[1mm]
$r=-1, 6, \frac{5}{2}\pm \frac{\sqrt{1-24(1+C)}}{2}$. & 
$r=-1, 0, 6, \mp\sqrt{1-48/C}$.\\[1mm] \hline 
\end{tabular} 
\vspace{5mm}

\subsection{The third level: find the constants of integration}

The third (and the last) level of the Painlev\'e test is a
 substitution into the original (not simplified) system $(7)$ the
 following series:
$$
z=\tilde a_1(t-t_0)^{-3}+\sum\limits_{j=1}^{r_{max}}
\widetilde{cz}_{j-3}(t-t_0)^{j-3}\quad\mbox{and}\quad
y=-\:\frac{15}{8}(t-t_0)^{-2}+
\sum\limits_{j=1}^{r_{max}}
\widetilde{cy}_{j-2}(t-t_0)^{j-2},
$$
where $r_{max}=6$,
$\widetilde{cy}_j$ and $\widetilde{cz}_j$ are unknown constants.

After this substitution system $(2)$ is transformed
to sequence of systems of linear algebraic equations. Solving these systems
we find $\widetilde{cy}_j$ and $\widetilde{cz}_j$.
Determinants of systems, which correspond to resonances, have to be zero.

For example, to determine $\widetilde{cy}_2$ and $\widetilde{cz}_1$ we
have obtained the following system:
$$
\left\{
\begin{array}{l@{}}
  557056\tilde a_1^6 + \tilde a_1^4\Big(15552000\lambda - 4860000\Big) 
+ \\[1mm]
 \tilde a_1^2\Big(864000000\widetilde{cy}_2 + 108000000\lambda^2 - 
67500000\lambda +10546875\Big)=0,\\[2mm] 818176\tilde a_1^4  + \tilde 
a_1^2\Big(15660000\lambda - 4893750\Big) - 810000000\widetilde{cy}_2- 
6328125=0.  \end{array} \right.  \eqno (11) $$

As one can see this system does not include $\widetilde{cz}_1$,
so $\widetilde{cz}_1$ is an arbitrary parameter (a constant of integration).
For any $\lambda$ this system can be solved as a system in
$\widetilde{cy}_2$ and $\tilde a_1$. We obtain new constant of integration
$\widetilde{cz}_1$, but we must fix $\tilde a_1$, so number of constants of
integration is equal to 2.  It is easy to verify that $\widetilde{cy}_4$ is
an arbitrary parameter, because the corresponding system is equivalent to
one linear equation.  So, we obtain an asymptotic solution which depends 
on three parameters, namely  $t_0$, \ $\widetilde{cz}_1$ and 
$\widetilde{cy}_4$.

\subsection{New asymptotic solutions}

Now it is easy to obtain asymptotic solutions with arbitrary accuracy.
For given $\lambda$ one has to choose $\tilde a_1$ as one of the roots
of system~$(11)$. After this the coefficients
$\widetilde{cz}_j$ and $\widetilde{cy}_j$ as functions of
$\widetilde{cz}_4$ and $\widetilde{cy}_6$ can be found automatically due to
computer algebra system {\bf REDUCE}. For some values
of $\lambda$ asymptotic solutions have been found as the following series
(without loss the generality we can put $t_0=0$):
$$
 z=\tilde a_1t^{-3}+\sum\limits_{j=1}^{50} \widetilde{cz}_{j-3}t^{j-3}
 \quad\mbox{and}\quad
y =-\:\frac{15}{8}t^{-2}+\sum\limits_{j=1}^{50}
\widetilde{cy}_{j-2}t^{j-2}.
$$

For example, if $\lambda=\frac{1}{9}$ then system $(11)$ has the following
solutions ($\tilde a_1\neq 0$):
$$
 \left\{\tilde a_1=\pm\:\frac{25\sqrt{2}}{16},   \quad
  \widetilde{cy}_2=-\:\frac{1819}{663552}\right\}, \qquad
 \left\{\tilde  a_1=\pm\:\frac{25i\sqrt{13}}{8\sqrt{374}},  \quad
\widetilde{cy}_2=-\:\frac{8700683}{1364926464}\right\}.
$$

If $\tilde a_1=\pm\frac{25\sqrt{2}}{16}$ then we obtain:  
$$ 
\begin{array}{r@{}c@{}l@{}} 
z\:&=&\:\frac{25\sqrt{2}}{16}t^{-3}+\frac{125}{192}t^{-2}+
\frac{25\sqrt{2}}{768}t^{-1}+\frac{1625}{82944}+\widetilde{cz}_1t+ \\[2mm]
&+&
\left(\frac{21845}{47775744}-\frac{\sqrt{2}}{6}\widetilde{cz}_1\right)t^2+
 \left(\frac{437425\sqrt{2}}{9172942848}-
\frac{25\sqrt{2}}{48}\widetilde{cy}_4 -
\frac{191}{3456}\widetilde{cz}_1\right)t^3+\dots,\\
\end{array}
$$
\vspace{-2.7mm}
$$
\mbox{ \ } \eqno(12.1)
$$
\vspace{-2.7mm}
$$
\begin{array}{@{}r@{}c@{}l@{}}
y\:&=&\:-\:\frac{15}{8}t^{-2}+\frac{5\sqrt{2}}{32}t^{-1}-\frac{205}{2304}+
 \frac{115\sqrt{2}}{13824}t- \frac{1819}{663552}t^2+ \\[2mm]
&+&
\left(\frac{1673\sqrt{2}}{11943936}+\frac{1}{6}\widetilde{cz}_1\right)t^3+
\widetilde{cy}_4t^4+\left(\frac{1044461\sqrt{2}}{220150628352}-
\frac{19}{9216}\widetilde{cz}_1-\frac{1}{2}\widetilde{cy}_4\right)t^5+
\dots
\end{array}
$$
and
$$
\begin{array}{r@{}c@{}l@{}}
z\:&=&\:-\:\frac{25\sqrt{2}}{16}t^{-3}+\frac{125}{192}t^{-2}-
\frac{25\sqrt{2}}{768}t^{-1}+\frac{1625}{82944}+\widetilde{cz}_1t+ \\[2mm]
&+&
\left(\frac{21845}{47775744}+\frac{\sqrt{2}}{6}\widetilde{cz}_1\right)t^2+
 \left(\frac{437425\sqrt{2}}{9172942848}-
\frac{25\sqrt{2}}{48}\widetilde{cy}_4 -
\frac{191}{3456}\widetilde{cz}_1\right)t^3+\dots,\\
\end{array}
$$
\vspace{-2.7mm}
$$
\mbox{ \ \  } \quad \eqno(12.2)
$$
\vspace{-2.7mm}
$$
\begin{array}{@{}r@{}c@{}l@{}}
y\:&=&\:-\:\frac{15}{8}t^{-2}-\frac{5\sqrt{2}}{32}t^{-1}-\frac{205}{2304}-
\frac{115\sqrt{2}}{13824}t-\frac{1819}{663552}t^2- \\[2mm]
&-&
\left(\frac{1673\sqrt{2}}{11943936}+\frac{1}{6}\widetilde{cz}_1\right)t^3+
\widetilde{cy}_4t^4-\left(\frac{1044461\sqrt{2}}{220150628352}-
\frac{19}{9216}\widetilde{cz}_1-\frac{23\sqrt{2}}{384}
\widetilde{cy}_4\right)t^5+
\dots
\end{array}
$$
For real values of parameters and time these solutions are real. 
In the case $a_1=\frac{25\sqrt{2}}{16}$ the 
following table shows how $\widetilde{cz}_{50}$ and $\widetilde{cy}_{50}$ 
 (coefficients of terms proportional to $t^{50}$) depend on the arbitrary 
parameters $\widetilde{cz}_{1}$ and $\widetilde{cy}_{4}$:

\begin{center}
\begin{tabular}{||c|c|c|c||c|c|c|c||}
\hline
$\widetilde{cz}_{1}$ & $\widetilde{cy}_{4}$&
$\widetilde{cz}_{50}$ & $\widetilde{cy}_{50}$&
$\widetilde{cz}_{1}$ & 
$\widetilde{cy}_{4}$ &
$\widetilde{cz}_{50}$ & $\widetilde{cy}_{50}$\\
\hline $-1$ & $-1$ & $4\!\times\!\!10^{-12}$& $- 1\!\!\times\!\!10^{-13}$&
0    &    0  & $ - 1\!\!\times\!\!10^{-44}$&
 $2\!\!\times\!\!10^{-45}$  \\ \hline
$-1$ & $-0.6$ & $4\!\!\times\!\!10^{-12}$& $5\!\!\times\!\!10^{-14}$&
0    &   0.2  &$ - 8\!\!\times\!\!10^{-20}$& $-3\!\!\times\!\!10^{-20}$\\
\hline
$-1$ & $-0.2$ & $-1\!\!\times\!\!10^{-17}$& $-2\!\!\times\!\!10^{-18}$&
0    &   0.4& $  - 2\!\!\times\!\!10^{-17}$&$ -
 9\!\!\times\!\!10^{-18}$\\
\hline
$-1$ & $0$ & $-1\!\!\times\!\!10^{-20}$& $3\!\!\times\!\!10^{-22}$&
0    &   0.6& $  - 5\!\!\times\!\!10^{-16}$&$ -
 5\!\!\times\!\!10^{-16}$\\
\hline
$-1$ & $0.4$ & $6\!\!\times\!\!10^{-15}$& $1\!\!\times\!\!10^{-16}$&
0    &  0.8& $  - 5\!\!\times\!\!10^{-15}$&$ -
 2\!\!\times\!\!10^{-15}$\\
\hline
$-1$ & $1$ & $6\!\!\times\!\!10^{-12}$& $8\!\!\times\!\!10^{-14}$&
0    &   1 & $    - 3\!\!\times\!\!10^{-14}$&$ -
 1\!\!\times\!\!10^{-14}$\\
\hline
$\!-0.6$ & $-1$ & $3\!\!\times\!\!10^{-12}$& $-5\!\!\times\!\!10^{-14}$&
0.4  &   0 & $  - 4\!\!\times\!\!10^{-25}$&$ -
 1\!\!\times\!\!10^{-26}$\\
\hline
$\!-0.6$ & $-0.6$ & $4\!\!\times\!\!10^{-14}$& $-1\!\!\times\!\!10^{-15}$&
0.4  &  0.4 &$    - 1\!\!\times\!\!10^{-15}$&$
 2\!\!\times\!\!10^{-17}$\\
\hline
$\!-0.6$ & $0$ & $-5\!\!\times\!\!10^{-23}$& $9\!\!\times\!\!10^{-25}$&
0.4  &  0.8 &$    - 3\!\!\times\!\!10^{-13}$&$
 2\!\!\times\!\!10^{-15}$\\
\hline
$\!-0.6$ & $0.4$ & $3\!\!\times\!\!10^{-15}$& $4\!\!\times\!\!10^{-17}$&
0.8  &  0 &$     - 1\!\!\times\!\!10^{-21}$&$
 -3\!\!\times\!\!10^{-23}$\\
\hline
$\!-0.6$ & $1$ & $3\!\!\times\!\!10^{-12}$& $2\!\!\times\!\!10^{-14}$&
0.8  &  0.4& $    - 2\!\!\times\!\!10^{-15}$&$
 2\!\!\times\!\!10^{-16}$\\
\hline
$\!-0.2$ & $-1$ & $1\!\!\times\!\!10^{-12}$& $-2\!\!\times\!\!10^{-14}$&
0.8  &   0.8& $   - 6\!\!\times\!\!10^{-13}$& $ 1\!\!\times\!\!10^{-14}$\\
\hline
$0$ & $-1$ & $-3\!\!\times\!\!10^{-14}$& $-1\!\!\times\!\!10^{-14}$&
1    &   1 & $ - 4\!\!\times\!\!10^{-12}$ & $  1\!\!\times\!\!10^{-13}$
\\ \hline
20    &  20 & $     - 2.2$&$   0.051$
& 20    &  40 &$      - 603$&$   6.88$\\
\hline
 40   &   20 &$    - 11.1 $& $ 0.01$&
 40   &   40 &     -  1128 & 24.5\\
\hline
\end{tabular}
\end{center}
One can see that coefficients tend to zero very rapidly when
the absolute values of the parameters are less than unit.

\section{The connection between asymptotic solutions and exact
            solutions}

E.I.~Timoshkova~\cite{Timosh} found that
solutions of the following equation
$$
  y_t^2=\tilde {\cal A} y^3+\tilde {\cal G} y^{5/2}+\tilde {\cal B} y^2+
  \tilde {\cal E} y^{3/2}+\tilde {\cal C} y+\tilde {\cal D}, \eqno (13)
$$
where $\tilde {\cal A}=\frac{-32}{15}$, $\tilde {\cal D}=0$, $\tilde {\cal
B}$, $\tilde {\cal C}$, $\tilde {\cal G}$ and $\tilde {\cal E}$  are some
depending on $\lambda$ constants, satisfy~$(3)$ at $C=-\:\frac{16}{5}$.  
After change of variables:  $y=\varrho^2$, we obtain the following 
equation:  
   $$ \varrho_t^2=\frac{1}{4}\Bigl(\tilde {\cal A}\varrho^4+ 
\tilde {\cal G} \varrho^3+\tilde {\cal B} \varrho^2 + \tilde {\cal 
   E}\varrho+\tilde {\cal C}\Bigr).  \eqno (14) 
   $$

Equation~$(14)$ is of P-type. The general solution
of equation~$(14)$ has only one arbitrary parameter $t_0$ and
can be expressed in elliptic functions~\cite{BE,Gera}.

For all values of $\lambda$ the trajectories of motion are given by the
following equation:
$$
  x^2\equiv z=-\:\frac{1}{2}\left({5\over 2}\tilde {\cal G} y^{3/2}+2(\tilde
{\cal B}+1)y+ \frac{3}{2}\tilde {\cal E} y^{1/2}+\tilde {\cal C}\right).
\eqno (15)
$$

Let us compare the asymptotic solutions~$(12.1)$ and~$(12.2)$ with these 
exact solutions.  It is easy to verify, for example, with the help of the 
Painlev\'e test of equation $(14)$, that at points of singularities these 
solutions and our asymptotic solutions have the same asymptotic behavior. 
It means that for some values of parameters $\widetilde{cz}_{1}$ and 
$\widetilde{cy}_{4}$ our asymptotic series give exact solutions and,
hence, converge.

Let us consider in detail the case of $\lambda=\frac{1}{9}$.
In this case the following solution has been found~\cite{Timosh}:
$$
  x^2\equiv z=-\:\frac{y}{3}\left\{\frac{5}{3}-
2i\sqrt{\frac{5y}{3}}\right\},
$$
where y is a solution of the following equation:
$$ 
y_t^2+\:\frac{32}{15}y^3+\frac{4}{9}y^2\pm\frac{8i}
  {\sqrt{135}}y^{5/2}=0. 
\eqno (16) 
$$ 

Depending on a choice of a sign before the last term, we obtain 
 either (in case of sign~$+$):
$$ 
  y=-\:\frac{5}{3\left(1-3\sin\left(\frac{t-t_0}{3} 
\right)\right)^{2^{\vphantom{27}}}} \qquad\mbox{and}\qquad 
  x^2=\frac{25\Big(1-\sin\left(\frac{t-t_0}{3}\right)\Big)}
{9\left(1-3\sin\left(\frac{t-t_0}{3}\right)\right)^{3^{\vphantom{27}}}};
\eqno (17.1)
$$
or (in case of  sign $ - $):
$$ 
  y=-\:\frac{5}{3\left(1+3\sin\left(\frac{t-t_0}{3} 
\right)\right)^{2^{\vphantom{27}}}}  \qquad\mbox{and}\qquad 
  x^2=\frac{25\Big(1+\sin\left(\frac{t-t_0}{3}\right)\Big)}
{9\left(1+3\sin\left(\frac{t-t_0}{3}\right)\right)^{3^{\vphantom{27}}}}.
\eqno (17.2)
$$

It is easy to verify that the series $(12.1)$ with 
$$ 
 \widetilde{cy_4}=-\:\frac{858455}{12039487488} \qquad\mbox{and}\qquad 
 \widetilde{cz_1}=\frac{3205\sqrt{2}}{3981312}, 
$$ 
are the Laurent series of $(17.1)$ and the series $(12.2)$ with 
$$ 
 \widetilde{cy_4}=-\:\frac{858455}{12039487488} \qquad\mbox{and}\qquad 
 \widetilde{cz_1}=-\:\frac{3205\sqrt{2}}{3981312}, 
$$ 
are the Laurent series of $(17.2)$.
So, our asymptotic series converge at these values of
parameters.  It is possible that our asymptotic series converge also for
other values of parameters. 
I plan to analyze this question in future.

\section*{CONCLUSIONS}

Using the Painlev\'e analysis one can not only find  
integrable cases of dynamical systems, but also construct 
special asymptotic solutions even in nonintegrable cases.

We have found the special solutions of the H\'enon--Heiles system
with $C=-\:\frac{16}{5}$ as formal Laurent series, depending on three 
parameters. For some values of two parameters the obtained solutions 
coincide with the known exact solutions. 

  The author is grateful to \  R.~I.~Bogdanov \ and \  V.~F.~Edneral \
for valuable discussions and \ E.~I.~Timoshkova \  for
comprehensive commentary of~\cite{Timosh}. This work has been supported by
the Russian Foundation for Basic Research under grants  {\the\textfont2 
N}\lower-0.4ex\hbox{\small\underline{$\circ$}}~00-15-96560   and 
{\the\textfont2  N}\lower-0.4ex\hbox{\small\underline{$\circ$}}~00-15-96577.

\pagebreak
\end{document}